# Intelligent Vehicle-Trust Point: Reward based Intelligent Vehicle Communication using Blockchain

Madhusudan Singh
Yonsei Institute of Convergence Technology
Yonsei University
Seoul, South Korea
msingh@yonsei.ac.kr

Shiho Kim
School of Integrated Technology,
Yonsei University
Seoul, South Korea
shiho@yonsei.ac.kr

*Abstract*—The Intelligent vehicle (IV) is experiencing revolutionary growth in research and industry, but it still suffers from many security vulnerabilities. Traditional security methods are incapable to provide secure IV communication. The major issues in IV communication, are trust, data accuracy and reliability of communication data in the communication channel. Blockchain technology works for the crypto currency, Bit-coin, which is recently used to build trust and reliability in peer-to-peer networks having similar topologies as IV Communication. In this paper, we are proposing, Intelligent Vehicle-Trust Point (IV-TP) mechanism for IV communication among IVs using Blockchain technology. The IVs communicated data provides security and reliability using our proposed IV-TP. Our IV-TP mechanism provides trustworthiness for vehicles behavior, and vehicles legal and illegal action. Our proposal presents a reward based system, an exchange of some IV-TP among IVs, during successful communication. For the data management of the IV-TP, we are using blockchain technology in the intelligent transportation system (ITS), which stores all IV-TP details of every vehicle and is accessed ubiquitously by IVs. In this paper, we evaluate our proposal with the help of intersection use case scenario for intelligent vehicles communication.

## I. Introduction (*Heading 1*)

Current ITS system uses ad-hoc networks for Vehicle communication such as DSRC, WAVE, Cellular Network, which does not guarantee secure data transmission. Currently, vehicle communication application security protocols are based on cellular and IT standard security mechanism which are not up-to-date and suitable for ITS applications. Still many researchers are working to provide standard security mechanism for ITS. Our proposed mechanism is advantages as it is easy to implement, it's a peer -to -peer communication, it provides a secure and trust environment for Vehicle communication with immutable database and ubiquitous data access in a secure way. Our proposal is based on a very simple concept of using Blockchain based trust environment for data sharing among Intelligent Vehicles using the IV-TP (Intelligent Vehicle-Trust Point). We are exploiting the features of Blockchain i.e. distributed and open ledger which is encrypted with Merkel tree and Hash function (SHA-256) and are based on Consensus Mechanism (Proof of Work Algorithm). We have not mentioned the details of the Blockchain mechanism for our application Intelligent Vehicle data sharing due to the limitation of space.

Previously, some researchers combined automotive and blockchain technology but most of them considered applications based on services and smart contracts. However, our proposal concentrates on secure and fast communication between intelligent vehicles (self-driving cars) [5]. We have proposed a unique crypto Intelligent Vehicle-Trust Point (IV-TP) based on blockchain technology. Our proposal explains the management of IV-TP for intelligent vehicles communication and we elicited the benefits of IV-TP and evaluated our proposal with intersection use case scenario for intelligent vehicles communication.

We organize our articles as follows; Section II presents the motivation of using Blockchain based trust environment for data sharing among Intelligent Vehicles using the IV-TP (Intelligent Vehicle-Trust Point) over traditional security methods. Section III presents the introduction of blockchain technology and existing work of blockchain technology for Intelligent Vehicles communication. Section IV, describes, our proposed reward based intelligent vehicles communication mechanism based on blockchain technology, Section V, discusses the generation of the intelligent vehicle trust point (IV-TP), Section VI evaluates our proposal with intersection based use case scenarios; Section VII concludes our paper, and discuss our future work for our proposed mechanism

## II. Related Work

### A. Blockchain Technology

Blockchain technology is distributed, open ledger, saved by each node in the network, which is self-maintained by each node. It provides peer-to-peer network without the interference of the third party. The blockchain integrity is based on strong cryptography that validates and chain blocks together on transactions, making it nearly impossible to tamper with any individual transaction without being detected [6].

Fig.4 shows the Blockchain technology features such as shared ledger, Cryptography, Signed blocks of transactions, and digital signatures [6].

This work was supported by Institute for Information & communications Technology Promotion(IITP) grant funded by the Korea government(MSIP) (No.2017-0-00560, Development of a Blockchain based Secure Decentralized Trust network for intelligent vehicles)

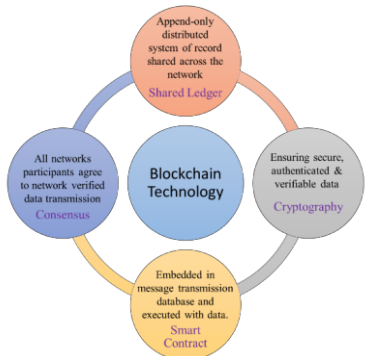

Fig. 1. Blockchain technology

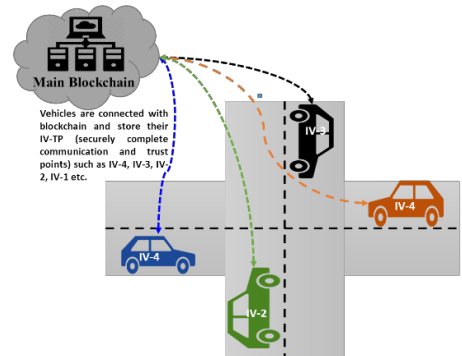

Fig. 2. Proposed blockchain Intelligent Vehicle Communication

*B. Previous work: Blockchain technology for Intelligent Transportation System.*

Yong yuan, et.al [7] has proposed the blockchain technology for ITS for establishment of secured, trusted and decentralized autonomous ecosystem and proposed a seven-layer conceptual model for the blockchain.

Benjamin et.al [8], have also proposed the blockchain technology for vehicular ad-hoc network (VANET). They have combined Ethereum' blockchain based smart contracts system with vehicle ad-hoc network. They have proposed combination of two applications, mandatory applications (traffic regulation, vehicle tax, vehicle insurance) and optional applications (applications which provides information and updates on traffic jams and weather forecasts) of vehicles. They have tried to connect the blockchain with VANET services. Blockchain can use multiple other functionalities such as communication between vehicles, provide security, provide peer-to-peer communication without disclosing personal information etc. Ali dorri et.al. [9] have proposed the blockchain technology mechanism without disclosing any private information of vehicles user to provide and update the wireless remote software and other emerging vehicles services. Sean Rowen et.al. [10] have described the blockchain technology for securing intelligent vehicles communication through the visible light and acoustic side channels. They have verified their proposed mechanism through a new session cryptographic key, leveraging both side-channels and blockchain public key infrastructure.

We define our blockchain mechanism for the intelligent vehicles communication environment. We propose the secure environment peer-to-peer communication between intelligent vehicles without interfering/disturbing other intelligent vehicles. We also evaluate our proposed mechanism with intersection road scenario based use case.

### III. INTELLIGENT VEHICLE-TRUST POINT: REWARD-BASED INTELLIGENT VEHICLES COMMUNICATION USING BLOCKCHAIN

We propose a reward based intelligent vehicles communication using blockchain technology. Our proposed mechanism has three basics technologies including communication network enabled connected device, Vehicular Cloud Computing (VCC) and blockchain technology (BT).

*A. Network enabled Connected device*

It is an internet-enabled device, which can organize, communicate in VANET such as Smartphone, PDA, Intelligent Vehicles, etc.

*B. Vehicular Cloud Computing*

VCC is a hybrid technology that has a remarkable impact on traffic management and road safety by instantly using vehicle resources, such as computing, data storage, and internet decision-making.

*C. Blockchain supported intelligent vehicles*

Blockchain consists of a technically unlimited number of blocks which are chained together cryptographically in chronological order. In this, each block consists of transactions, which are the actual data to be stored in the chain.

### IV. VEHICLES -TRUST POINT GENERATION

We propose an Intelligent Vehicles -Trust Point (IV-TP) crypto unique ID that is issued by vehicle seller/authorized dealers. This IV-TP is developed by blockchain crypto mechanism and is similar to bitcoin. This IV-TP is issued to every intelligent vehicle. During communication, vehicles provide IV-TP to build trust in the communication network. The Vehicular networks having blockchain enabled service/user data providers, manages the IV-TP.

IV-TP is an encrypted unique number, which uniquely issued to every IV and called as IV-TP ID. Every IV has its own IV-TP ID, generated by the authorized authority. The IV-TP is earned by calculating some computation in the group communication. Greater the IV-TP attained by an IV, higher will be its respect and honor. With the help of IV-TP, one can get the complete history of vehicles (accident history, condition of IV, crime history, etc.). The IV-TP access method is show in figure 6.

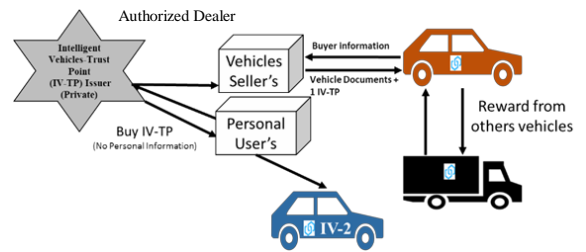

Fig. 3. IV-TP access methods

Blockchain technology based intelligent vehicles communicate with each other following the steps shown below: We have explained the process of IV-TP sharing and verification in figure 7.

### A. Key generation

Firstly, each IV will generate its private and public key. The blockchain will maintain the public key of all IVs in network and when an IV want to communicate another IV then it will access the public of another IV from the blockchain.

### B. Digital Signature:

Secondly, each vehicle shall digitally sign the message to check integrity and non-repudiation of the message. With digitally signed message, receiver can easily find, that the message is not tampered, and the sender of the message is a valid IV in the network.

### C. Verification

Lastly, the receiver after receiving message identifies the sender by verifying the digitally signed encrypted message. After verification, receiver decrypts the message with the

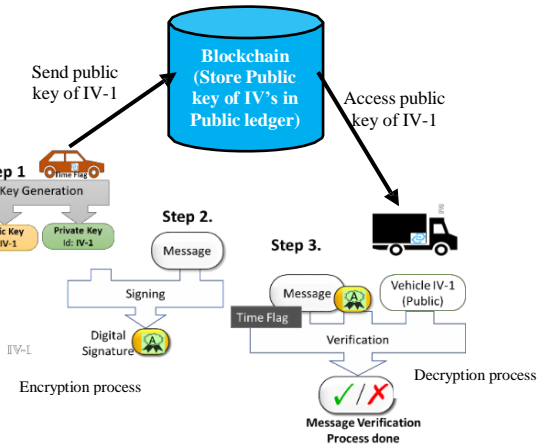

Fig. 4. Message process between two IV

- Example Setting of IV network

Consider, five intelligent vehicles having IV-TP as IV-1, IV-2, IV-3, IV-4, & IV-5, respectively. Message is broadcasted from IV-1 to IV-5 as shown in figure 8.

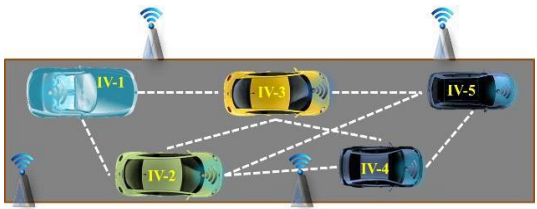

Fig. 5. Intelligent Vehicular networks

Each vehicle broadcasts the message in the network. The blockchain maintains a table shown in Table 1 showing the information of "who communicates with whom". It does not require any personal information of any IV. It only needs IV-TP of IVs.

TABLE 1. COMMUNICATION INFORMATION OF IVS ON BLOCKCHAIN

| Intelligent Vehicles | Communicated Vehicles |
|---|---|
| IV-1 | IV-2, IV-3, IV-4 |
| IV-2 | IV-3, IV-4, IV-1 |
| IV-2 | IV-4, IV-1, IV-2 |
| IV-4 | IV-1, IV-2, IV-3 |

Each IV broadcast the message in the network, using the message framework shown in fig. 9.

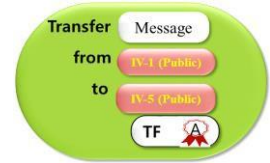

Fig. 6. Message Framework example

In fig 9, IV-1 and IV-5 are IV-TP of intelligent vehicles 1 and 5 respectively. TF is time flag of message broadcasted from intelligent vehicles.

### D. Consensus Protocols

Broadcasted message will be validated only after verification by more than 50% of the network vehicles. This validation process is based on Proof of Driving (PoD) algorithm. PoD provides evidence that the vehicles are legal and are running in the same network shared by the approved vehicles at the communication time. All vehicles communication data will be managed on the vehicular cloud with the IV-TP ID. If in future, IV owners want to sell or change their IVs, then they can access their complete data history via the vehicular cloud.

## V. USE CASE EVALUATION OF BLOCKCHAIN ENABLED IV COMMUNICATION

We evaluate our proposed method with the help of use case. We randomly select the intersection scenario as a use case example for the explanation of our proposal.

### A. IV communication on Intersection scenario

Consider, four IVs (IV-1, IV-2, IV-3, and IV-4) reach the intersection, almost at the same time, as shown in figure 10. In this condition, how our proposed system will help to overcome this situation. Before coming to the intersection, each IV will broadcast its status (a message that they want to cross the intersection) on the network. Every IV near the intersection will receive the broadcasted message in the network. They will first verify the IV-TP ID from VC, and then each vehicle will calculate the received time of the broadcasted message of the vehicles. The vehicle, which first calculates the time, will broadcast the vehicles IV-TP ID, which first arrived at the intersection. Other vehicles will also calculate and verify the first arrived vehicle at the intersection. If everyone agrees, they will give way to the first arrived vehicle to go first or cross the intersection first. In table 2, we show the IV-TP ID of each vehicle and their message broadcasted time, and received time from other vehicles.

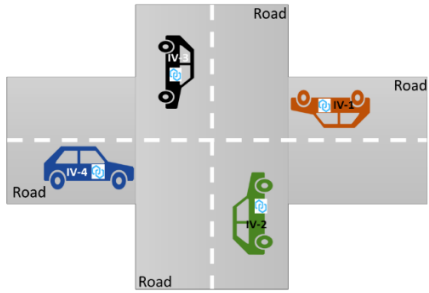

Fig. 7. Intersection scenario for IV communication

According to the table 2 shown below, IV-1 comes first on the Intersection region and IV-3 calculates first so, IV-3 will broadcast the message that IV-1 will move first based on first come first serve approach and IV-3 will also propose move sequences for other vehicles ( IV-1then next IV-2, then next IV-3 and last IV-4). Other IVs (IV-1, IV-2, and IV-4) will calculate the time and agree with the schedule given by IV-3. IV-1 will get some reward in terms of IV-TP with its own IV-TP ID. All this information of the vehicles will be stored on the vehicular cloud with their IV-TP IDs.

TABLE 2. TIME SCHEDULE OF MESSAGE BROADCASTED AND RECEIVED BY IV'S

| IV-TP | Message Transmission Time (sec.) | Message Received by IV-TP | Receiver Time (sec.) | Reward |
|---|---|---|---|---|
| IV-1 | 1.00 | IV-2 | 1.01 | |
| | | IV-3 | 1.03 | |
| | | IV-4 | 1.07 | |
| IV-2 | 1.01 | IV-1 | 1.00 | |
| | | IV-3 | 1.03 | |
| | | IV-4 | 1.07 | |
| IV-3 | 1.03 | IV-1 | 1.00 | IV-1->IV-3 0.5 bit trust |
| | | IV-2 | 1.01 | |
| | | IV-4 | 1.07 | |
| IV-4 | 1.07 | IV-1 | 1.00 | |
| | | IV-2 | 1.01 | |
| | | IV-3 | 1.06 | |

Note: our proposed mechanism use case is for general vehicles and not specified for special vehicles such as ambulance, police, VIP's vehicles, etc.

In our use case, the vehicles are an intelligent machine (internet connected self-driving vehicles) so, they have enough computational power to calculate time.

## VI. CONCLUSION

In this paper, we have presented a reward based intelligent vehicle communication based on blockchain technology and not for specific services as previously proposed by other researchers. We have proposed crypto IV-TP that will help to improve the privacy of IVs. IV-TP provide fast and secure communication between IVs. It also helps to detect the detailed history of IVs communication. IV communication data will be stored on the VC, as long as the user wants. During any accident, the IVs communication history and their reputations are ubiquitously available to authorized organizations (hospital, insurance company, police etc.) and home via VC.

In future, we will simulate our proposed mechanism for multiple vehicle communication scenarios as well as analyze different use cases (suspicious actions by IVs and managing IV-TP etc.) with a solution.

ACKNOWLEDGMENT

This work was supported by Institute for Information & communications Technology Promotion (IITP) grant funded by the Korea government (MSIT) (No.2017-0-00560, Development of a Blockchain based Secure Decentralized Trust network for intelligent vehicles).